\documentclass[
 reprint,
 amsmath,amssymb,
 aps
]{revtex4-1}

\usepackage{enumerate}
\usepackage[none]{hyphenat}
\usepackage{color}

\usepackage{siunitx}

\usepackage{array}
\usepackage{multirow}
\usepackage{booktabs}
\usepackage{adjustbox}
\usepackage{dcolumn}

\usepackage{graphicx}
\usepackage[font=small,labelfont=bf]{caption}
\usepackage{subcaption}
\usepackage{float}
\usepackage[section]{placeins}

\usepackage{hyperref} 
\usepackage{subfiles}

\hypersetup{
	colorlinks=true,
	citecolor=red,
	linktoc=all,
	linkcolor=blue,
	urlcolor=blue,
	linktocpage
}

\begin{document}

\title{Energy Confinement Time Scaling Law Derived from Paz-Soldan NF 2024}
\author{Priyansh Lunia}
 \email{p.lunia@columbia.edu}
\author{A.O. Nelson}
\author{C. Paz-Soldan}
\affiliation{%
 Applied Physics and Applied Mathematics, Columbia University\\
 New York, NY 10027
}

\begin{abstract}
\textbf{Abstract} --- Results from C. Paz-Soldan \textit{et al} 2024 \textit{Nucl. Fusion} \textbf{64} 094002 \cite{paz2024nt} demonstrate encouraging energy confinement properties for the negative triangularity scenario, similar to or exceeding the scaling of the IPB98(y,2) law. This work describes the procedure with which a new scaling law was regressed specifically from the data from the aforementioned article. Given the relatively small size of the single-machine dataset, kernel density estimation was employed to minimize sampling bias and bootstrapping was used to give a realistic estimate of the large uncertainties from the regression. The resulting power law shows a robustly stronger dependence on plasma current and more severe power degradation as compared to the H-mode scaling law.
\end{abstract}

\maketitle

\section{Introduction}
\label{sec:intro}
The negative triangularity (NT) scenario was further explored in dedicated studies in \cite{paz2024nt,thome2024overview}. These studies demonstrated the simultaneous attainment of $H_{98y2}>1$, $f_\mathrm{GW}>1$, and $\beta_N>2.5$. Of particular note for this work is the energy confinement performance of NT compared to the IPB98(y,2) scaling law \cite{transportChapter2Plasma1999}. The NT scenario showed H-mode-like confinement scaling with the robust avoidance of edge-localized modes, illustrating the potential of NT as a reactor-relevant scenario. 

\begin{figure}[h!]
  \centering
  \begin{subfigure}[b]{\linewidth}
  	\centering
    \includegraphics[width=0.9\linewidth]{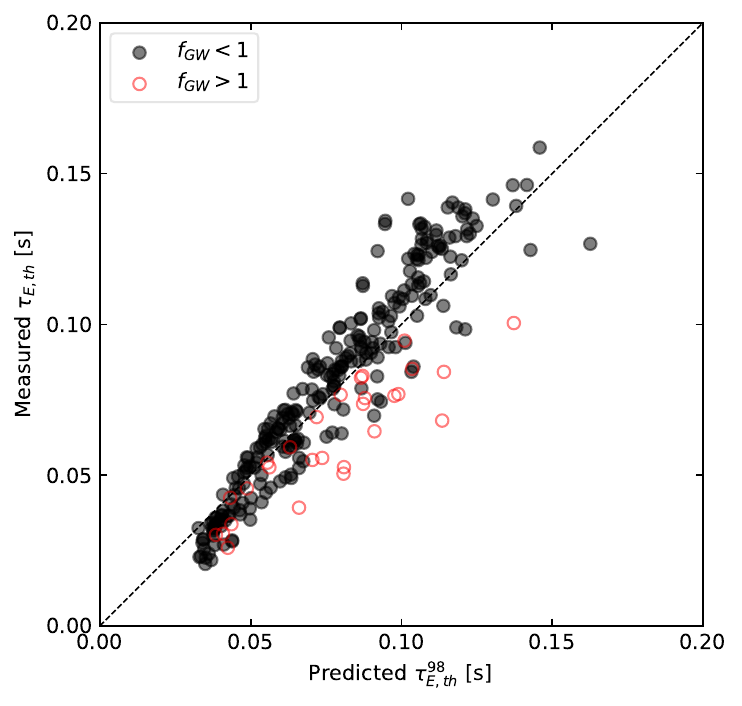}
    \caption{Dataset plotted against the $H_{98}$ scaling.}
    \label{fig:h98}
  \end{subfigure}
  \begin{subfigure}[b]{\linewidth}
  	\centering
    \includegraphics[width=0.9\linewidth]{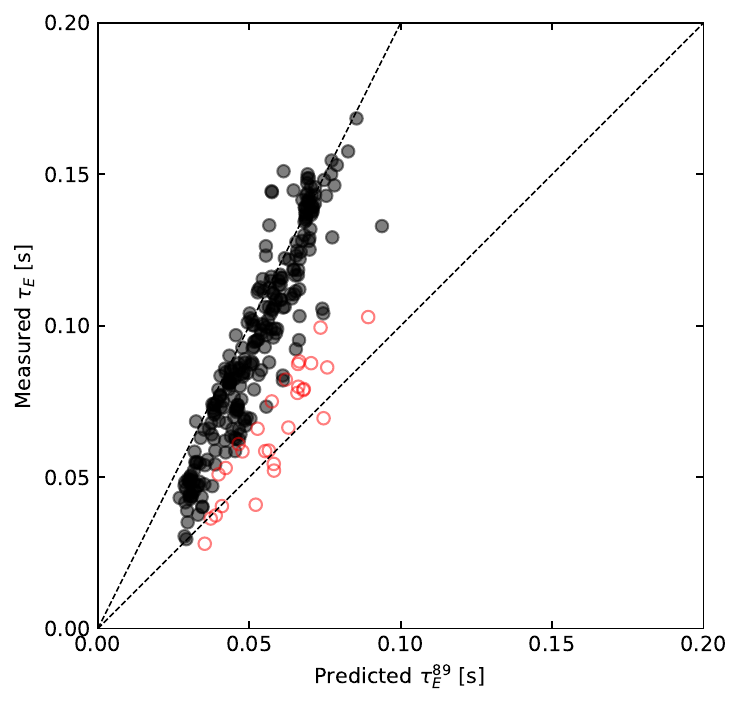}
    \caption{Dataset plotted against the $L_{89}$ scaling.}
    \label{fig:l89}
  \end{subfigure}
  \caption{Comparison of (a) $H_{98}$ scaling prediction against measured $\tau_{E,th}$, (b) $L_{89}$ scaling prediction against measured $\tau_E$. Points with $f_{GW}>1$ are highlighted in red.}
  \label{fig:scaling}
\end{figure}

The NT database in the supplementary material of Ref. \cite{paz2024nt} is utilized. Fig. \ref{fig:h98} shows the consistency of the dataset's (thermal) energy confinement time, $\tau_{E,th}$, against the IPB98(y,2) scaling law. It can be seen that NT largely follows H-mode-like scaling, with a divergence at high $\tau_{E,th}$. Additionally, Fig. \ref{fig:l89} shows a similar plot with the measured energy confinement time, $\tau_E$, against the prediction of the $L_{89}$ L-mode scaling law. Note that the distinction between $\tau_{E,th}$ and $\tau_E$ is that the contribution of the fast particles is subtracted from the former. This figure shows a clear deviation of the confinement time from the L-mode scaling, where higher $\tau_E$ approach $\sim 2\times$ the performance as predicted by the $L_{89}$ scaling law. The comparisons in Fig. \ref{fig:scaling} clearly demonstrate the different behavior of NT confinement time scaling compared to the standard H-mode and L-mode scaling laws, motivating the need for a new engineering scaling regression specifically for NT.

This report provides an extension to the regression results published in Ref. \cite{paz2024nt}, giving additional detail on the regression procedure and uncertainty quantification --- this is covered in the following section. Section \ref{sec:future} discusses potential future work to further elucidate the NT confinement time scaling law. The scripts used in this analysis are available on \href{https://github.com/hansec/ColumbiaMCF/tree/nt_scaling}{GitHub}.

\section{Scaling Law}
\label{sec:scale}
The usual power-law scaling approach is used for the energy confinement time regression. Specifically, we start with the following form:
\begin{equation}
\label{eq:power}
	\tau_{E,th} = CI_p^{\alpha_I}B_t^{\alpha_B}n_e^{\alpha_n}P_\mathrm{tot}^{\alpha_p}\kappa^{\alpha_\kappa}\epsilon^{\alpha_\epsilon}R_0^{\alpha_R}M^{\alpha_M},
\end{equation}
where we seek the values of the exponents, $\alpha$.
We then perform a log-linear transformation to Eq. \ref{eq:power} due to two reasons. Firstly, this transformation results in the errors in the fit being multiplicative instead of additive, which is appropriate for a power law. Secondly, the log-linear transform enables linear least squares to be utilized, which ensures the global minimum in the fit is found. The regression problem now looks like:
\begin{equation}
\label{eq:log_power}
\begin{split}
	\log{\tau_{E,th}} &= \log{C} 
	\\ &+ \alpha_I \log{I_p} + \alpha_B \log{B_t} + \alpha_n \log{n_e} + \alpha_P \log{P_\mathrm{tot}} 
	\\& + \alpha_\kappa \log{\kappa} + \alpha_\epsilon \log{\epsilon} + \alpha_R\log{R_0} + \alpha_M\log{M}.
\end{split}
\end{equation}

\begin{figure}
  \centering
  \includegraphics[width=0.9\linewidth]{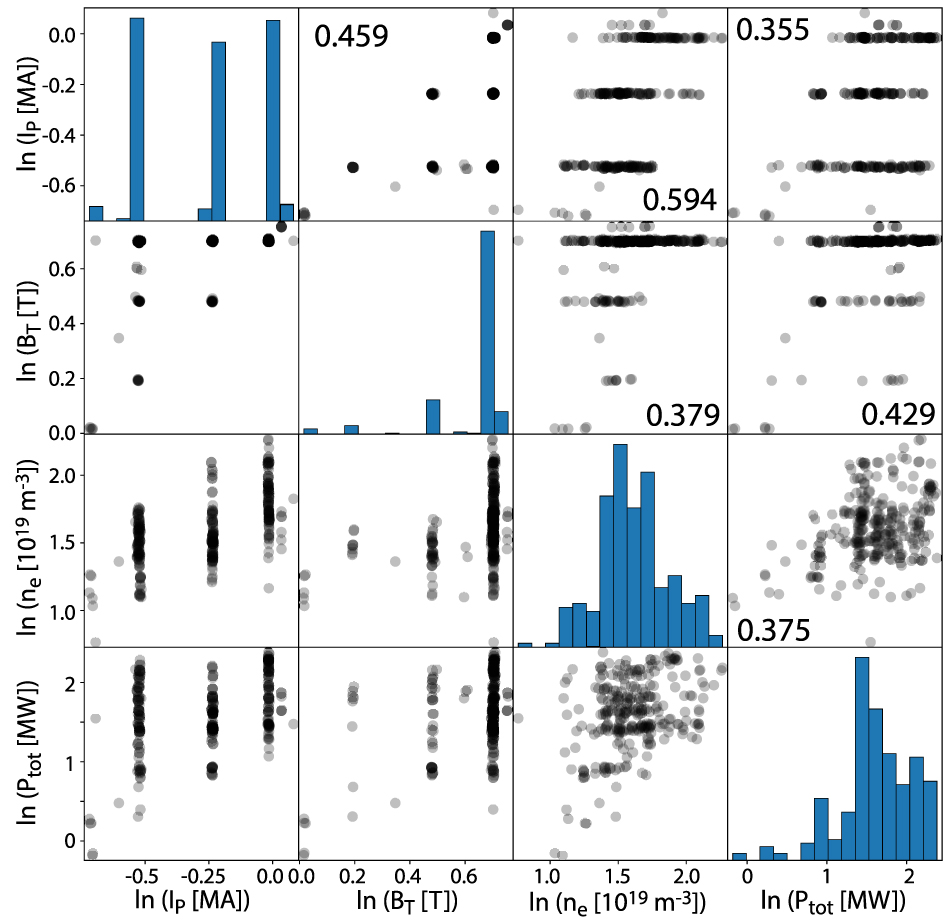}
  \caption{Data scatter matrix for all data used in the regression, reproduced from Appendix A of Ref. \cite{paz2024nt}. The Pearson correlation coefficient is indicated in the corresponding subplots, and does not show significant co-linearity.}
  \label{fig:data}
\end{figure}

As previously mentioned, the NT database in the supplementary material of Ref. \cite{paz2024nt} is utilized, with additional filters applied. Specifically, stationary phases with too high $\dot{W}$, too high saturated $n=2$ mode activity, $f_{GW}>1$, and zero torque are omitted. This results in 260 stationary phases to fit to, across 132 discharges. Given all the data in the dataset is from a single device with minor changes in the shaping parameters ($\kappa$, $\epsilon$, $R_0$), it is not possible to fit to these parameters. Additionally, there was no variation in the mass of the main ion species, $M$, in the whole dataset. As a result, the exponents for these parameters are fixed to those from the IPB98(y,2) scaling law. The fitting data is displayed in Fig. \ref{fig:data}, reproduced from Appendix A of Ref. \cite{paz2024nt}. 

\begin{figure}
  \centering
  \includegraphics[width=0.9\linewidth]{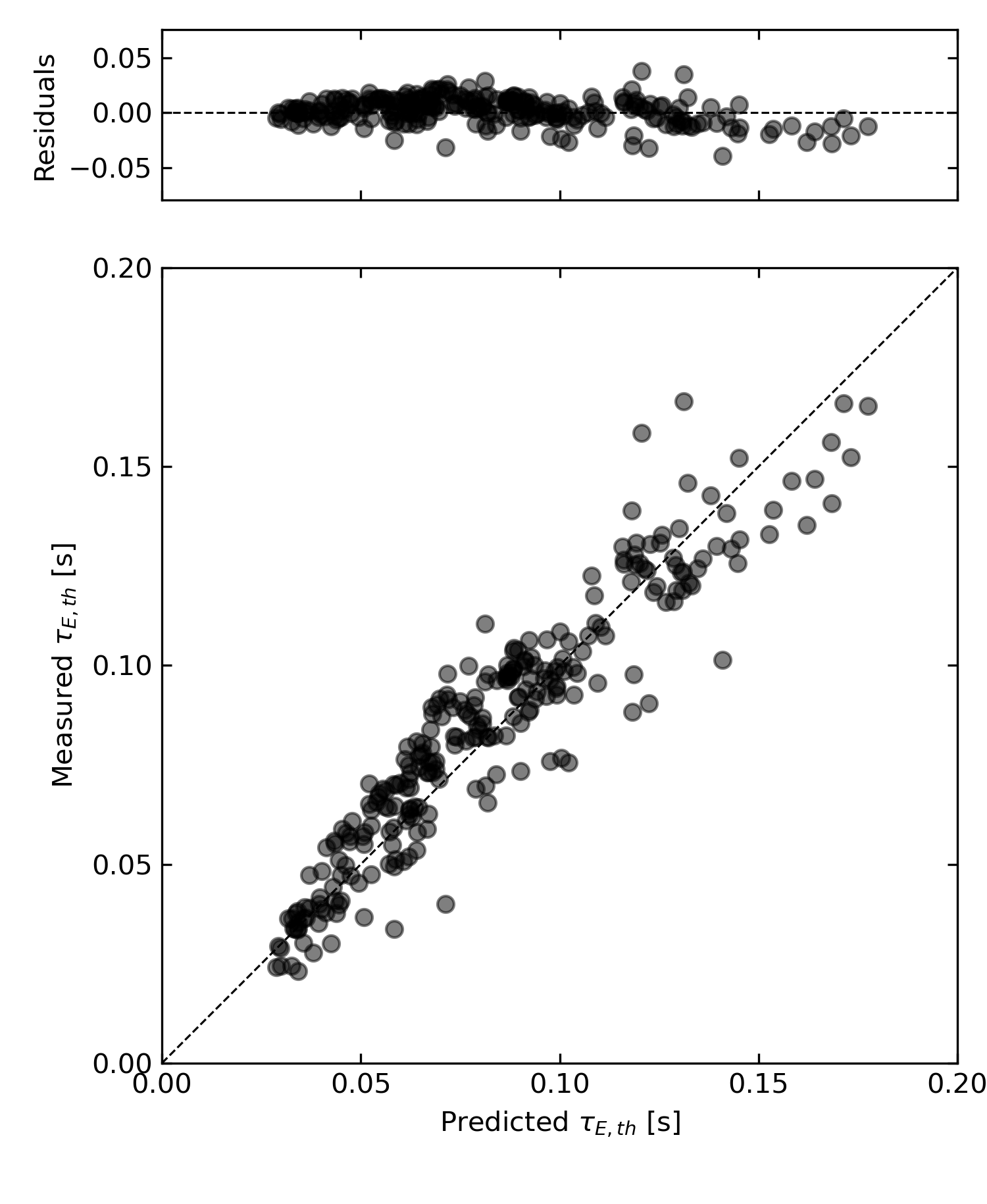}
  \caption{KDE-weighted fit to Eq. \ref{eq:log_power}, with the residuals shown in the top plot. This fit obtains an $R^2$ score of 0.877.}
  \label{fig:kde}
\end{figure}

It can be seen in Fig. \ref{fig:data} that a uniform distribution does not exist across all the parameters, which would lead to linear least squares having sampling bias. To combat this, kernel density estimation (KDE) is used to assign weights to the data points such that less sampled regions of the space are weighted more strongly and more sampled regions of the space are weighted more weakly. This technique has been used in  error field penetration scaling studies to alleviate sampling bias \cite{logan2020robustness}. The regression to Eq. \ref{eq:log_power} can be seen in Fig. \ref{fig:kde}. 

\begin{figure}
  \centering
  \includegraphics[width=0.9\linewidth]{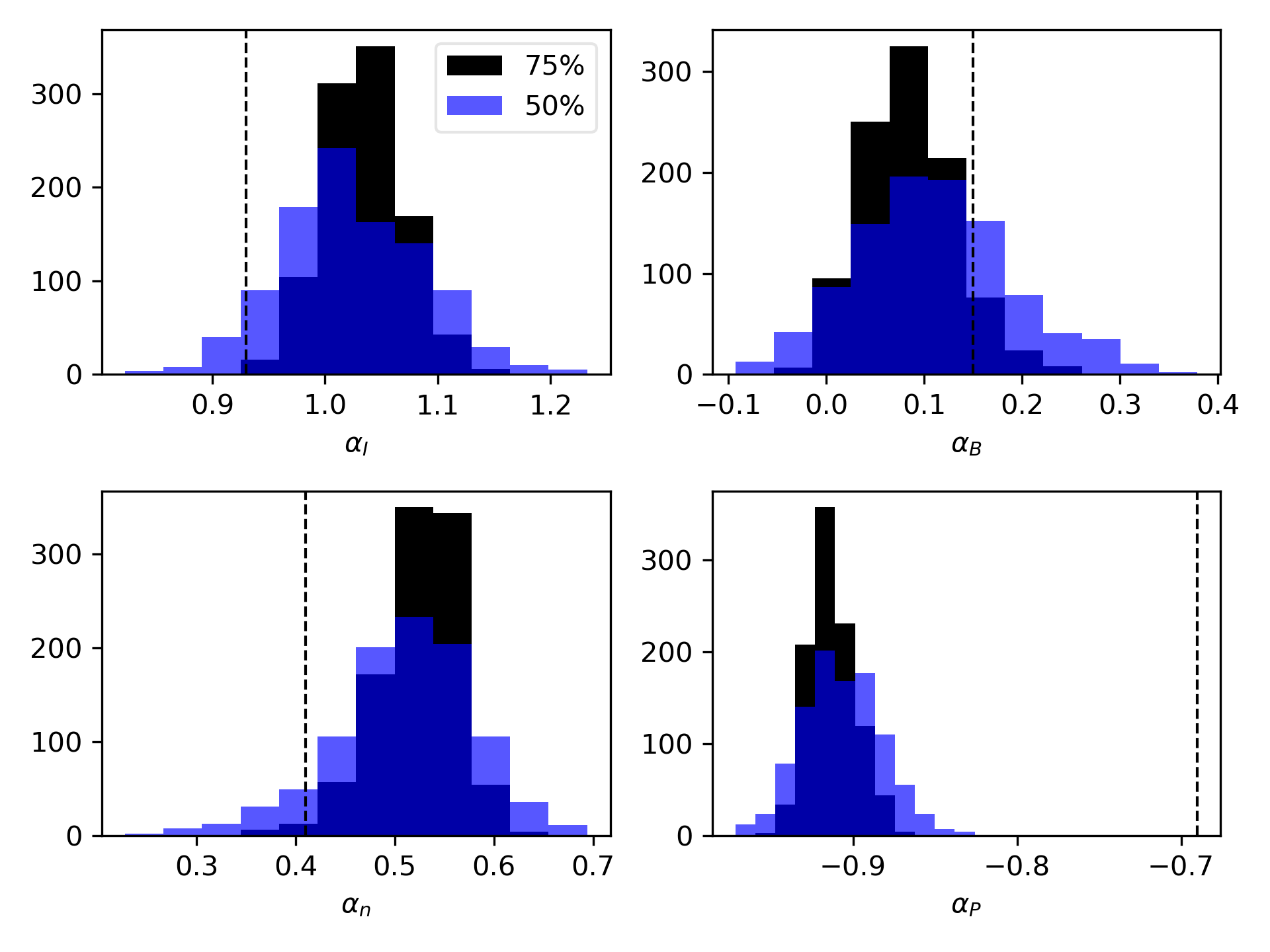}
  \caption{Bootstrapped KDE-weighted fits of 75\% and 50\% randomly-sampled subsets of the full dataset. Dashed lines indicate the values from the IPB98(y,2) scaling law.}
  \label{fig:bootstrap}
\end{figure}

It is important to note here the importance of uncertainty quantification, especially so for this particular regression as it only utilizes data from a single machine. Even large multi-machine datasets can have significant uncertainties, which has been shown in recently updated $\tau_{E,th}$ H-mode scaling laws \cite{verdoolaege2021scaling}. 

A simple yet effective method for quantifying the uncertainties in this regression is by bootstrapping, which involves performing numerous regressions on a randomly sampled subset of the full dataset. Fig. \ref{fig:bootstrap} illustrates this technique on 1000 regressions of 75\% and 50\% randomly sampled subsets. It is clear that there is a large spread in the values of the exponents given the choice of data, but clear trends can be seen in the exponents compared to those from IPB98(y,2). Namely, there is a stronger dependence on $I_p$ and $P_\mathrm{tot}$. To capture the largest reasonable uncertainties, we tentatively take the mean of each quantity's 50\% resampled distribution plus/minus $2\sigma$, obtaining a 95\% confidence interval. The results of the regression and uncertainty quantification are summarized in Table \ref{tbl:exp}.\\
\begin{table}
\begin{ruledtabular}
    \begin{tabular}{ldd}
      	\textrm{Quantity} & \textrm{Value} & \textrm{Uncertainty} \\
      	\colrule\colrule
      	Coefficient & 0.0821 & 0.018\\
      	$I_p$ & 1.02 & 0.12 \\
	    $B_t$ & 0.11 & 0.16 \\
      	$n_e$ & 0.51 & 0.14 \\
      	$P_\mathrm{tot}$ & -0.91 & 0.05 \\
      \bottomrule 
    \end{tabular}
    \caption{\label{tbl:exp}Exponent values and uncertainties given by the mean and $2\sigma$, respectively, of the 50\% resampled distributions in Fig. \ref{fig:bootstrap}. The value for the coefficient does not correspond to an exponent, but rather the coefficient, $C$, itself.}
\end{ruledtabular}
\end{table}

\section{Future Work}
\label{sec:future}

Given that the regression results presented in the previous section were only using data from a single device, scaling with the shaping parameters ($\kappa$, $\epsilon$, and $R_0$) was not possible due to their low variance \cite{paz2024nt}. As such, a natural extension would be to include NT data from additional machines that expands the dataset in the operating space of these shaping parameters. This may introduce co-linearities given the correlation between $\epsilon$ and $R_0$, which could be addressed using the ridge regression technique.

This work has exclusively performed weighted linear least squares regression, which, while straightforward, may not be the robust technique to use. Indeed, the work in Ref. \cite{verdoolaege2021scaling} has taken extensive effort to robustly regress a large multi-machine dataset using Bayesian techniques which properly addresses sampling bias, co-linearities, and uncertainty quantification. Implementing this technique for NT scaling would be a natural next step.


\bibliography{bibliography.bib} 

\end{document}